\documentclass[aps,showpacs,floatfix,preprint,superscriptaddress]{revtex4}
\usepackage{epsfig}

\newcommand{\half}{\frac{1}{2}}
\newcommand{\ihsp}{\hspace*{\fill} }

\newcommand{\be}{\begin{equation}}
\newcommand{\ee}{\end{equation}}
\newcommand{\bea}{\begin{eqnarray}}
\newcommand{\eea}{\end{eqnarray}}

\newcommand{\DS}{Dyson--Schwinger }
\newcommand{\BS}{Bethe--Salpeter }
\newcommand{\WT}{Ward-Takahashi }

\newcommand{\Eq}[1]{Eq.~(\ref{#1})}
\newcommand{\Eqs}[1]{Eqs.~(\ref{#1})}
\newcommand{\Fig}[1]{Fig.~{\ref{#1}}}

\newcommand{\Ref}[1]{Ref.~\cite{#1}}
\newcommand{\Refs}[1]{Refs.~\cite{#1}}

\newcommand{\w}{\omega}
\newcommand{\G}{\Gamma}
\newcommand{\s}{\!\cdot\!}

\newcommand{\al}{\alpha}
\newcommand{\de}{\delta}
\newcommand{\si}{\sigma}
\newcommand{\ga}{\gamma}
\newcommand{\ro}{\rho}
\newcommand{\la}{\lambda}
\newcommand{\ov}[1]{\overline{#1}}
\newcommand{\dk}[1]{\,\,\,\raisebox{-0.4ex}{\large $\bar{}$}\!\!d\,{#1}\,}
\def\slr#1{\setbox0=\hbox{$#1$}           
   \dimen0=\wd0                                 
   \setbox1=\hbox{/} \dimen1=\wd1               
   \ifdim\dimen0>\dimen1                        
      \rlap{\hbox to \dimen0{\hfil/\hfil}}      
      #1                                        
   \else                                        
      \rlap{\hbox to \dimen1{\hfil$#1$\hfil}}   
      /                                         
   \fi}

\begin{document}
\title{Bethe--Salpeter Meson Masses Beyond Ladder Approximation}
\author{P.~Watson}
\email{peter.watson@theo.physik.uni-giessen.de}
\author{W.~Cassing}
\email{wolfgang.cassing@theo.physik.uni-giessen.de}
\affiliation{Institute for Theoretical Physics, University of Giessen, 
Heinrich-Buff-Ring 16, 35392 Giessen, Germany}
\author{P.~C.~Tandy}
\email{tandy@cnr2.kent.edu}
\affiliation{Center for Nuclear Research, Department of Physics,  Kent State 
University, Kent OH 44242, USA}
\date{June 30, 2004}

\begin{abstract}
The effect of quark-gluon vertex dressing on the ground state masses
of the $u/d$-quark  pseudoscalar, vector and axialvector mesons is 
considered with the Dyson-Schwinger equations.  This extends the 
ladder-rainbow \BS kernel  to 2-loops.  To render the
calculations feasible for this exploratory study, we employ a simple
infrared dominant model for the gluon exchange that implements the
vertex dressing.   The resulting model, involving two distinct 
representations of the effective gluon exchange kernel, preserves both
the axial-vector Ward-Takahashi identity and  charge conjugation symmetry.
Numerical results confirm that the  
pseudoscalar meson retains its Goldstone boson character.  The vector meson
mass, already at a very acceptable value at ladder level, receives only 
30~MeV of attraction from this vertex dressing.   For the axial-vector states,
which are about 300~MeV too low in ladder approximation, the results are
mixed:  the $1^{+-}$ state receives 290~MeV of repulsion, but the $1^{++}$ 
state is lowered further by 30~MeV.   The exotic channels $0^{--}$ and 
$1^{-+}$ are found to have no states below 1.5~GeV in this model.  

\end{abstract}
\preprint{Kent State U preprint no. KSUCNR-204-02}
\maketitle
\section{\label{sec:intro}Introduction}

In the modeling of QCD for hadron physics, the rainbow truncation of the 
quark propagator \DS equation (DSE) coupled with the ladder truncation of the 
\BS bound state equation (BSE) has been found to provide a very efficient 
description for ground states~\cite{Jain:1993qh,Maris:2003vk} and for 
finite temperature and density~\cite{Roberts:2000aa}.
In particular, a ladder-rainbow model~\cite{Maris:1999nt},  with one 
infrared parameter to generate the empirically acceptable amount of  
dynamical chiral symmetry breaking,  provides  an excellent description 
of the ground state pseudoscalars and vectors including the charge form 
factors~\cite{Maris:2000sk,Volmer:2000ek}, 
electroweak and strong decays~\cite{Maris:2001am,Jarecke:2002xd}, and 
electroweak transitions~\cite{Maris:2002mz,Ji:2001pj}.  Recent 
comprehensive reviews of QCD analysis and modeling for nonperturbative 
physics emphasize the gauge sector DSEs~\cite{Alkofer:2000wg} and hadron 
physics~\cite{Maris:2003vk}.   The ladder-rainbow truncation is known to 
satisfy both the vector  and the
flavor non-singlet axial-vector \WT identities; the latter implementation
of chiral symmetry guarantees the Goldstone boson nature of the flavor 
non-singlet 
pseudoscalars independently of model details~\cite{Maris:1998hd}.  The small
explicit symmetry breaking through current masses provides the detailed
description of the pseudoscalar masses.  

Although the vector masses are not explicitly protected by a symmetry, 
the excellent
description in ladder-rainbow truncation (typically within 5\% of 
experiment~\cite{Maris:1999nt}) illustrates the strong correlation between 
hyperfine splitting amongst ``S-wave'' states and  dynamical chiral 
symmetry breaking.
However, the ladder-rainbow truncation has inadequacies that are beginning 
to be understood and addressed.  Corrections to the bare quark-gluon 
vertex inherent in the ladder-rainbow kernel have been examined within
a schematic infrared dominant model~\cite{Munczek:1983dx} that admits 
algebraic analysis.  In a dressed loop expansion, the corrections
to the ladder-rainbow \BS kernel were found to have repulsive and attractive
terms that almost completely cancel for pseudoscalars and vectors but not
so for scalars~\cite{Bender:1996bb}.   Our understanding is far from 
complete; the schematic model used for this analysis does not bind scalar or
axial vector meson states.

In the axial vector channels \mbox{$1^{++}$} ($a_1/f_1$) and 
\mbox{$1^{+-}$} ($b_1/h_1$), the ladder-rainbow models, exemplified by 
\Refs{Maris:1999nt,Alkofer:2002bp},
are too attractive: the masses produced are 0.8-0.9~GeV~\cite{MarisPrivCom}. 
Evidently the orbital excitation energy \mbox{$m_{a_1} - m_\rho \approx$}  
is about a factor of 3 too small.   On the other hand, very reasonable 
$m_{a_1}$ and $m_{b_1}$ values ($\approx$ 1.3~GeV) are obtained with 
covariant separable models~\cite{Burden:1997nh,Burden:2002ps,Bloch:1999vk} 
that incorporate some of the key features of the ladder-rainbow truncation.  
This encourages a systematic examination.

The ladder \BS kernel is vector-vector coupling 
($\gamma_\mu \otimes \gamma_\nu$) and this generates a particular 
coupling of quark spin and orbital angular momentum.  Some of the 
processes beyond this level constitute dressing of the quark-gluon vertex
which generates a more general Dirac matrix structure for the
vertex.  Although  12 independent covariants are needed to describe
the most general dressed vertex, those that involve the scalar Dirac
matrix are of particular interest.  Meson bound states are dominated
by the infrared and one expects that it is most important to model
the dressed vertex at very low gluon momentum.  In this case only the
quark momentum and the Dirac matrices are available for construction
of the vector vertex covariants; of the three possible covariants, 
two involve  
$\gamma_\mu$ and the third involves the Dirac scalar matrix.  The latter
generates a different coupling of quark spin and orbital angular momentum; 
such a correction to ladder-rainbow could distinguish between 
``P-wave'' states such as the axial-vectors and the ``S-wave'' 
pseudoscalars and vectors.   Another characteristic of a Dirac scalar
matrix term in the vertex is that, in the chiral limit, it cannot be
generated by any finite order of perturbation theory; it is generated by 
dynamical chiral symmetry breaking in quark propagators internal to
the dressed vertex.   Since dynamical chiral symmetry breaking 
sets the infrared scale of the ladder-rainbow truncation, it is natural 
to include the Dirac scalar part of the dressed vertex in considerations
beyond this level.

It is only recently that lattice-QCD has begun to provide information
on the infrared structure of the dressed quark-gluon 
vertex~\cite{Skullerud:2003qu}.  In the absence of well-motivated 
nonperturbative models for the vertex, many authors 
have employed the (Abelian) Ball-Chiu Ansatz~\cite{Ball:1980ay} times the 
appropriate color matrix and comprehensive results from a truncation of the 
coupled 
gluon-ghost-quark DSEs have been obtained this way~\cite{Fischer:2003rp}.  
However, there is no known way to develop a BSE kernel that is 
dynamically matched to such quark propagator solutions in the sense
that chiral symmetry is preserved through the axial-vector \WT identity.
It is known that such a symmetry imposes a specific dynamical relation
between the quark self-energy and the BSE kernel~\cite{Munczek:1995zz}.
Since the ladder-rainbow kernel contains infrared phenomenology matched
to the chiral condensate, a specification of correction terms that ignores
chiral symmetry  will needlessly alter the pseudoscalar sector.  An
artificial fine tuning of parameters can misrepresent the relationship
to other sectors. 

There is a known  constructive scheme~\cite{Bender:1996bb} that defines 
a diagrammatic expansion of 
the BSE kernel corresponding to any diagrammatic expansion of the quark 
self-energy such that the axial-vector \WT  identity is preserved.   
We will apply this to a 1-loop model of the dressed vertex.  Since use of 
a finite range effective gluon exchange kernel to construct the 2-loop BSE 
kernel leads to a very large computational
task, especially with retention of all possible covariants,  we begin here
with a simplification.  To an established finite range rainbow self-energy,
we add a 1-loop vertex dressing model using the Munczek-Nemirovsky 
(MN)~\cite{Munczek:1983dx} delta function model.   The corresponding 
BSE kernel is then formed by the previously mentioned constructive scheme,
with one modification: to preserve charge conjugation symmetry for 
appropriate meson
solutions, this approach requires that the BSE kernel be symmetrized 
with respect to interchange of the two distinct effective gluon exchange
models that appear therein.     The resulting kernel still preserves the 
flavor non-singlet axial-vector \WT identity.    This hybrid model retains
the advantages of a finite range ladder-rainbow term, providing about 0.9~GeV
towards axial-vector masses, while enabling a feasible exploration of the
effects of  vertex dressing.   Recent investigations of the 
BSE beyond rainbow-ladder truncation~\cite{Bender:2002as,Bhagwat:2004hn} 
have exploited the algebraic structure that follows from use of the MN model 
throughout and have not been able to address axial-vector states.

In Section~\ref{sec:vert} we consider vertex dressing within the quark DSE
and specify the employed rainbow self-energy
and the model dressed vertex.  The resulting 2-loop self-energy of the
present approach is described.  The chiral-symmetry-preserving 
BSE kernel is obtained from this self-energy in Section~\ref{sec:BSkernel}
where the preservation of the axial-vector \WT identity in the present 
context 
is outlined.  We discuss the solutions for the dressed quark propagator
in Section~\ref{sec:quark}.   In Section~\ref{sec:bse} we describe the
BSE of the present work.   Numerical results for meson masses
are presented in Section~\ref{sec:res}.   Section~\ref{sec:conc} contains
a summary and the Appendix provides details of the 2-loop quark DSE
that arises here.

\section{\label{sec:vert}Vertex Dressing and the quark \DS Equation}

We work with the Euclidean metric wherein
Hermitian Dirac $\ga$-matrices obey $\{\ga_{\mu},\ga_{\nu}\}=2\de_{\mu\nu}$, 
and scalar products of 4-vectors denote \mbox{$a\cdot b =$} 
\mbox{$ \sum_{i=1}^4 a_i b_i$}.   The color group is $SU(N_c)$ with $N_c=3$.  

In QCD, the \DS equation for the renormalized quark propagator is 
\be
S^{-1}(p)=Z_2(\imath\slr{p} + m) + Z_{\rm 1F}\, \int^\Lambda \dk{k}\, 
   g^2 D_{\mu\nu}(p-k) \,\frac{\la^a}{2}\ga_{\mu}\, S(k)\, 
                                        \frac{\la^a}{2}\G_{\nu}(k,p)~~~,
\label{eq:qdse0}
\ee
where $\dk{k}=d^4k/(2\pi)^4$, $g$ is the renormalized coupling constant,
the $\la^a/2$ are the $SU(3)$
color matrices, $m$ is the quark bare mass, $D_{\mu\nu}(q)$ is 
the renormalized dressed gluon propagator in Landau gauge ($q=p-k$), and $\G_{\nu}(k,p)$ 
is the renormalized dressed quark-gluon vertex.  Here $Z_{\rm 1F}$ and 
$Z_2$ are the vertex and quark field renormalization 
constants.   With a translationally invariant ultra-violet regularization of 
the integrals characterized by mass scale $\Lambda$, the 
renormalization conditions are $S(p)^{-1}=\imath\slr{p}+m_{\rm R}(\mu)$ and 
\mbox{$\G_{\nu}(p,p) = \ga_{\nu}$} at a sufficiently large spacelike 
renormalization point $p^2=\mu^2$.   Here $m_{\rm R}$ is the renormalized
current mass related to $m$ through \mbox{$Z_2\, m = Z_4\, m_{\rm R}$}, 
with $Z_4$ being the renormalization constant for the scalar component
of the self-energy.  

Most studies have used the rainbow truncation in which the kernel of 
\Eq{eq:qdse0} becomes
\begin{equation}
\label{rainbowansatz}
 g^2 D_{\mu \nu}(q)\, Z_{1{\rm F}}\, \frac{\la^a}{2} \Gamma_\nu(k,p) 
    \rightarrow \Delta_{\mu\nu}(q)\,\frac{\la^a}{2} \gamma_\nu \,,
\end{equation}
where \mbox{$\Delta_{\mu\nu}(q) =$} \mbox{$t_{\mu\nu}(q)\,\Delta(q^2)$}, 
$t_{\mu\nu}(q)$ is the transverse projector, and $\Delta(q^2)$ is an 
effective interaction.   Due to chiral symmetry, there is a close 
dynamical connection between the kernel of \Eq{eq:qdse0} and the
\BS kernel for pseudoscalar mesons~\cite{Bender:1996bb,Maris:1998hd}.
This connection, and the observation that $\Delta(q^2)$ should implement
the leading renormalization group scaling of the gluon propagator, the 
quark propagator and the vertex, has been exploited~\cite{Maris:1997tm} 
to specify the ultraviolet
behavior of $\Delta(q^2)$ by that of the renormalized quark-antiquark
interaction or ladder \BS kernel.  In such a QCD renormalization
group improved ladder-rainbow model~\cite{Maris:1997tm}, $\Delta(q^2)$ 
behaves as $4\, \pi\,\alpha_s^{\rm 1-loop}(q^2)/q^2$ in the ultraviolet 
and is parameterized in the infrared.
The resulting rainbow \DS equation reproduces the leading logarithmic
behavior of the quark mass function in the perturbative spacelike region.
The corresponding ladder \BS kernel is 
\be
K \rightarrow  -\, \Delta_{\mu\nu}(q)\, \frac{\la^a}{2}\ga_{\mu}\,\otimes
        \frac{\la^a}{2}\ga_{\nu}~~~.
\label{eq:KBSE}
\ee

Independently of the details of the model, this
ladder-rainbow truncation preserves chiral symmetry as expressed in the
axial-vector \WT identity, and thus guarantees massless pseudoscalars in 
the chiral limit~\cite{Maris:1998hd}.   It is known that the chiral 
symmetry relation between the kernels of the \DS and \BS equations may 
be maintained order by order beyond the ladder-rainbow level in a 
constructive scheme in which the first two terms are~\cite{Bender:1996bb}
\be
Z_{\rm 1F}\, \frac{\la^a}{2}\G_{\nu}(k,p)= \frac{\la^a}{2}\ga_{\nu} - 
   \int\dk{l}\, \ov{\Delta}_{\ro\la}(l)\, \frac{\la^b}{2}\ga_{\ro}\,
S(k+l) \,\frac{\la^a}{2}\ga_{\nu}\, S(p+l) \,\frac{\la^b}{2}\ga_{\la}~~~.
\label{eq:qgv0}
\ee
This replaces the corresponding portion of the \DS kernel of 
\Eq{rainbowansatz}.   Normally one has \mbox{$\ov{\Delta}=$} 
\mbox{$\Delta$}.

The corresponding \BS kernel in this scheme is generated  as the sum of 
terms produced by cutting a quark line in the diagrammatic representation 
of the quark self-energy.
Due to the complexity of a \BS calculation with a two-loop
kernel, existing numerical 
implementations~\cite{Bender:1996bb,Bender:2002as,Bhagwat:2004hn} for 
bound states have used the infrared dominant model in which 
\mbox{$\Delta(q^2) =$} 
\mbox{$\ov{\Delta}(q^2) \propto $} \mbox{$\delta^4(q)$}.   Through its
results for dressed quark propagators and pseudoscalar and vector mesons, 
it is known that this simple model effectively summarizes the
qualitative behavior of
more realistic models.   The strong infrared enhancement is the dominant
common feature.   The algebraic structure that this model gives 
to the DSE-BSE system was exploited to implement the BSE kernel 
consistent with an Abelian-like summed ladder model for the quark-gluon 
vertex~\cite{Bender:2002as}.   It was found that the quark propagator and 
the BSE meson masses are well-represented by the 1-loop version of that 
vertex model.  A disadvantage of this algebraic model for the entire
BSE kernel is that it does not support non-zero relative momentum for 
quark and anti-quark.  This is likely the reason why it does not bind
the  (P-wave) axial-vector states.

Since existing finite range ladder-rainbow models typically produce
axial vector masses of about 0.9~GeV, we use this to build a model 
that generalizes the approach of \Ref{Bender:1996bb};
the delta function interaction is used only as an effective representation
of the gluon exchange that implements vertex dressing via \Eq{eq:qgv0}.   
Since this hybrid model has \mbox{$\ov{\Delta} \neq \Delta$}, an adjustment
must be made in the construction of the symmetry-preserving \BS kernel and
we address this in Section~\ref{sec:BSkernel}.
  
Hence for $\Delta_{\mu\nu}(q)$ appropriate to the ladder-rainbow level of
the present approach, we take the following convenient 
representation~\cite{Alkofer:2002bp}:
\be
\Delta_{\mu\nu}(q)=4\pi^2\, D\, t_{\mu\nu}(q)\, \frac{q^2}{\w^2}\, 
                                  \exp{\left(-\frac{q^2}{\w^2}\right)}~~~.
\label{eq:FRgluon}
\ee
The parameters $D$ and $\w$ are required to fit the light pseudoscalar 
meson data and 
the chiral condensate; throughout we will use the values~\cite{Alkofer:2002bp} 
$\w=0.5$~GeV, $D=16~{\rm GeV}^{-2}$.  We shall consider only $u/d$-quarks with 
$m=5$~MeV representing the current mass~\cite{Alkofer:2002bp}.
With this form we retain the phenomenological successes of recent
studies~\cite{Maris:1997tm,Maris:1999nt,Alkofer:2002bp} of the light quark 
flavor non-singlet mesons at the rainbow-ladder level.
We note that the above model does not implement the ultraviolet behavior
of the QCD running coupling; this contributes typically  10\% or less to 
meson masses~\cite{Maris:1999nt} and this level of precision is not
our concern here.   The form of this model interaction also produces
ultraviolet convergent integrals and thus renormalization will be 
unnecessary here; one has \mbox{$Z_{\rm 1F} = Z_2 = 1$} and 
\mbox{$m_{\rm R} = m$}. 

\begin{figure}[t]
\mbox{\epsfig{figure=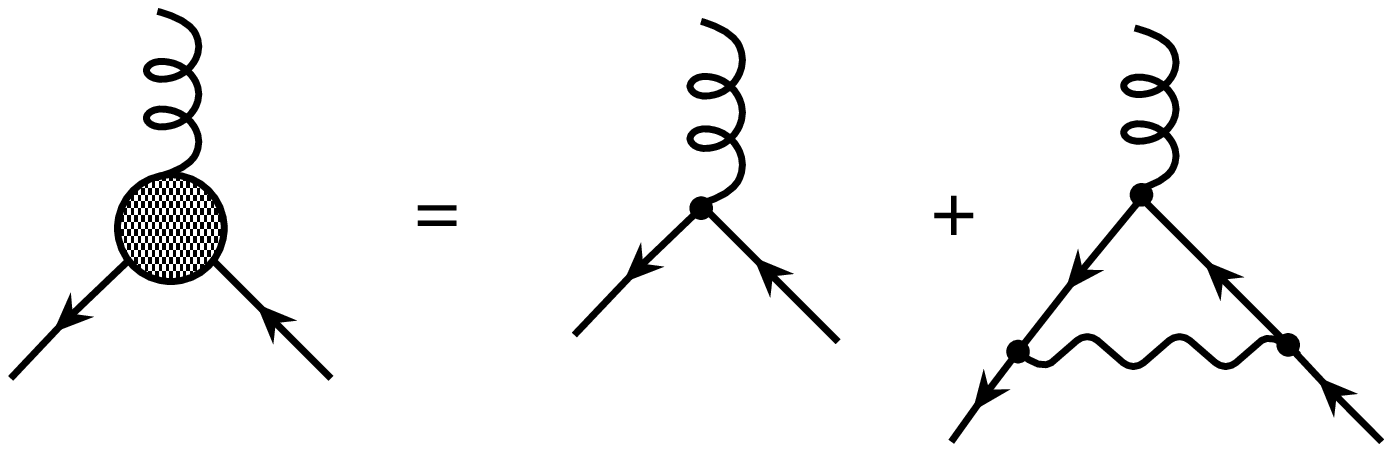,width=10cm}}
\caption{\label{fig:vert}The approximation employed here for the dressed 
quark-gluon vertex.  To provide for manageable calculations, the internal 
gluon line that implements the dressing is simplified  to the 
$\de$-function form $\ov{\Delta}$ given in \protect\Eq{MNgluon}, and 
identified in diagrams as a wavy line.}
\end{figure}

For the effective gluon exchange  that implements vertex dressing, we employ 
the Munczek-Nemirovsky (MN) model~\cite{Munczek:1983dx} in the form 
\be
\ov{\Delta}_{\ro\la}(l)=(2\pi)^4\, G\, \de^4(l)\, t_{\ro\la}(l)~~~.
\label{MNgluon}
\ee
The single 
parameter $G$ represents the integrated infrared strength which is the 
key feature for empirically successful ladder-rainbow models.   To set 
$G$, one could match the quark
mass function at \mbox{$p^2 = 0$} or the vector meson ground state mass
in rainbow-ladder truncation.    Here we 
have chosen to simply demand that both models have the same integrated
strength.  From \Eqs{eq:FRgluon} and (\ref{MNgluon}), this  gives 
$G=D\,\w^4/2$, and yields $G=0.5~{\rm GeV}^2$.  We note that the 
smaller estimate 
\mbox{$G \approx 0.25~{\rm GeV}^2$} is suggested by reproduction of 
\mbox{$m_\rho =$} \mbox{$ 0.770~{\rm GeV}$} in rainbow-ladder
truncation.  We present our results for the parameter range 
\mbox{$G = 0-0.5~{\rm GeV}^2$}. 

The simple form, \Eq{MNgluon}, for $\ov{\Delta}$ reduces \Eq{eq:qgv0} for the 
dressed vertex to an algebraic form.  With 
\mbox{$\la^b/2\, \la^a/2\, \la^b/2 = $} \mbox{$(-1/6)\, \la^a/2$},
one obtains
\be
\G_{\nu}(k,p)=\ga_{\nu} +
       \frac{G}{8}\,\ga_{\ro}\,S(k)\,\ga_{\nu}\,S(p)\,\ga_{\ro}~~~,
\label{eq:qgv1}
\ee
where the factor $1/8$ comes from the above $1/6$ times an extra factor
of $3/4$ generated by the combination of the transverse projector and
the $\delta$ function.   With this, \Eq{eq:qdse0} for the quark propagator
becomes 
\be
S^{-1}(p)=\imath\slr{p} + m + \int\dk{k}\Delta_{\mu\nu}(p-k)
        \left\{ \frac{4}{3}\,\ga_{\mu}\,S(k)\,\ga_{\nu} + 
        \frac{G}{6}\,\ga_{\mu}\,S(k)\,\ga_{\ro}\,S(k)\,\ga_{\nu}
        \,S(p)\,\ga_{\ro} \right\}~~~.
\label{eq:qdse}
\ee

Our approximation for the dressed quark-gluon vertex is summarized by 
\Fig{fig:vert} where the wavy line identifies the gluon exchange 
that has been simplified to the MN model. 
We note that, although in an Abelian theory like QED this is the only
1-loop  diagram that provides vertex dressing, this is not the case in QCD
where there is a second 1-loop quark-gluon diagram allowed due the 
existence of a 3-gluon vertex.     
There is no definitive information available for nonperturbative 
modeling of the 3-gluon vertex and such considerations are beyond the 
scope of this work.   We note that in \Ref{Bhagwat:2004hn} the effect
of such a contribution was estimated by arguing that a rescaling
of the Abelian-like 1-loop diagram for the quark-gluon vertex would be the 
result.   The examination of \BS bound states in this context again
utilized the algebraic structure afforded by the
MN model.  We explore the present model for its capability
to addresss a larger class of meson states (e.g., axial-vectors) through 
an extension of 
\Ref{Bender:1996bb} where the vertex phenomenology is based upon \Eq{eq:qgv0}.
The establishment of this capability may allow a wider examination of
vertex phenomenology in future investigations. 

\section{\label{sec:BSkernel}Symmetry-preserving \BS Kernel}

The close dynamical connection between the \BS kernel and
the quark propagator is manifest in chiral symmetry as expressed through 
the axial-vector Ward-Takahashi identity.   We shall be interested only
in the color singlet, flavor non-singlet channels
where such an identity leads to the Goldstone phenomenon in the chiral limit.
The flavor singlet channels have an axial anomaly term 
in the axial-vector Ward-Takahashi identity, which blocks the Goldstone 
phenomenon; the \mbox{$\eta-\eta^\prime$} system and scalars with flavor 
singlet
components lie outside of our considerations.    From the quark self-energy
of our  model as given in \Eq{eq:qdse}, the symmetry-preserving
BSE kernel can be obtained by the constructive scheme of 
\Ref{Bender:1996bb} with a generalization to account for the concurrent use
of two distinct effective interactions.  

In a flavor non-singlet channel, and with equal mass quarks, the 
axial-vector Ward-Takahashi identity is
\be
- \imath P_{\mu}\,\G_{\mu}^5(p;P) = S^{-1}(p_+)\ga_5+\ga_5S^{-1}(p_-)
                                            - 2m_{\rm R}\,\G^5(p;P)~~~,
\label{AV_WTI}
\ee
where we have factored out the explicit flavor matrix.  The color-singlet 
quantities 
$\G_{\mu}^5$ and $\G^5$ are the axial-vector vertex and the 
pseudoscalar vertex respectively, $p$ is the relative quark-antiquark 
momentum, and $P$ is the total momentum.  We use the notation $p_+=p+\xi P$ 
and $p_-=p-(1-\xi)P$, where the momentum sharing parameter $\xi=[0,1]$ 
represents the freedom of choice of relative momentum in
terms of individual momenta.  Due to Poincar\'e invariance, results will 
be independent of $\xi$ if a complete momentum and Dirac matrix 
representation is used~\cite{Maris:2000sk}.   In the present work 
with equal mass quarks, we use $\xi=1/2$ for convenience.

The BSE kernel determines the dressed parts
of the vertices $\G_{\mu}^5$ and $\G^5$, while the self-energy determines
the dressed part of $S^{-1}$.    Corresponding to a given 
approximation for $S^{-1}$, one seeks the matching approximation for 
the BSE kernel so that the above \WT identity holds for the truncated
theory.  In this way the pseudoscalar bound state results will be 
dictated by the pattern of chiral symmetry breaking and largely invariant 
to model details.  

If we define
\be
\Lambda(p;P) = - \imath P_{\mu}\,\G_{\mu}^5(p;P) + 2m_{\rm R}\,\G^5(p;P)~~~,
\label{Lambda}
\ee
then the BSE integral equations for $\G_{\mu}^5$ and $\G^5$, that have
inhomogeneous terms $Z_2\,\ga_5\ga_\mu$ and $Z_4\,\ga_5$ respectively, can
be combined to yield
\be
\Lambda(p;P)_{\al\beta}=
         [-\imath Z_2\,\ga_5\slr{P}+Z_4\,2m_{\rm R}\ga_5]_{\al\beta} +
\int^\Lambda\dk{k}K(p,k;P)_{\al\beta,\de\ga}[S(k_+)\Lambda(k;P)
                                                  S(k_-)]_{\ga\de}~~~,
\label{eq:ldse0}
\ee
where $K(p,k;P)_{\al\beta,\de\ga}$ is the BSE kernel.  For generality and 
clarity we have retained the renormalization constants although, in the
eventual application to the present model, they will be unity.
If \Eq{AV_WTI}
is to hold, then we may eliminate $\Lambda(p;P)$ in favor of $S^{-1}$
so that \Eq{eq:ldse0} explicitly relates the BSE kernel and the dressed quark 
propagator via
\bea
[S^{-1}(p_+)\ga_5+\ga_5S^{-1}(p_-)]_{\al\beta} &=& 
    Z_2[-\imath\ga_5\slr{P} + 2m\ga_5]_{\al\beta}\nonumber\\
 && + \int^\Lambda\dk{k}K(p,k;P)_{\al\beta,\de\ga}[\ga_5S(k_-)
  + S(k_+)\ga_5]_{\ga\de}~~~.
\eea
It is helpful to move the inhomogeneous term to the left hand side so
that cancellations leave only the self-energy integrals on the left.  
With use of the explicit form of the self-energy integrals from the
DSE, as given by \Eq{eq:qdse0}, we thus have the axial-vector 
\WT identity expressed in the equivalent form 
\bea
Z_{\rm 1F}\frac{4}{3}\int^\Lambda\dk{k}&&\!\!\!\!\!\!\left\{
     g^2D_{\mu\nu}(p_+-k)[\ga_{\mu}S(k)\G_{\nu}(k,p_+)\ga_5]_{\al\beta} 
+ g^2D_{\mu\nu}(p_--k)[\ga_5\ga_{\mu}S(k)\G_{\nu}(k,p_-)]_{\al\beta}\right\}
\nonumber\\&&
=\int^\Lambda\dk{k}K(p,k;P)_{\al\beta,\de\ga}[\ga_5S(k_-)
                                     +S(k_+)\ga_5]_{\ga\de}~~~.
\label{eq:axwi1}
\eea
For a given approximation or truncation to the 
vertex $\G_{\nu}(k,p)$, the corresponding truncated BSE kernel 
has to satisfy this integral relation to preserve chiral symmetry.  
Any proposed Ansatz  can be checked through substitution.  The
rainbow truncation on the left (substitute \Eq{rainbowansatz}) and 
the ladder truncation on the right (substitute \Eq{eq:KBSE})
obviously satisfy \Eq{eq:axwi1}. 

The general relation between the BSE kernel $K$ and the quark 
self-energy $\Sigma$ can be expressed through the functional 
derivative~\cite{Munczek:1995zz} 
\be
K(x^\prime, y^\prime; x, y) = - \frac{\delta}{\delta S(x,y)} \,
                                  \Sigma(x^\prime, y^\prime)~~~.
\label{Kernelderiv}
\ee
It is to be understood that this procedure is defined in the presence 
of a bilocal external source for $\bar q q$ and thus $S$ and $\Sigma$ are
not translationally invariant until the source is set to zero after 
the differentiation.  An appropriate formulation is the 
Cornwall-Jackiw-Tomboulis effective 
action~\cite{Cornwall:1974vz}.  In this context, the above 
coordinate space formulation ensures the correct number of independent 
space-time variables will be manifest.    
Fourier transformation of that 4-point function to momentum 
representation produces $K(p,q;P)$ having the correct momentum 
flow appropriate to the BSE kernel for total momentum $P$.   

\begin{figure}[t]
\mbox{\epsfig{figure=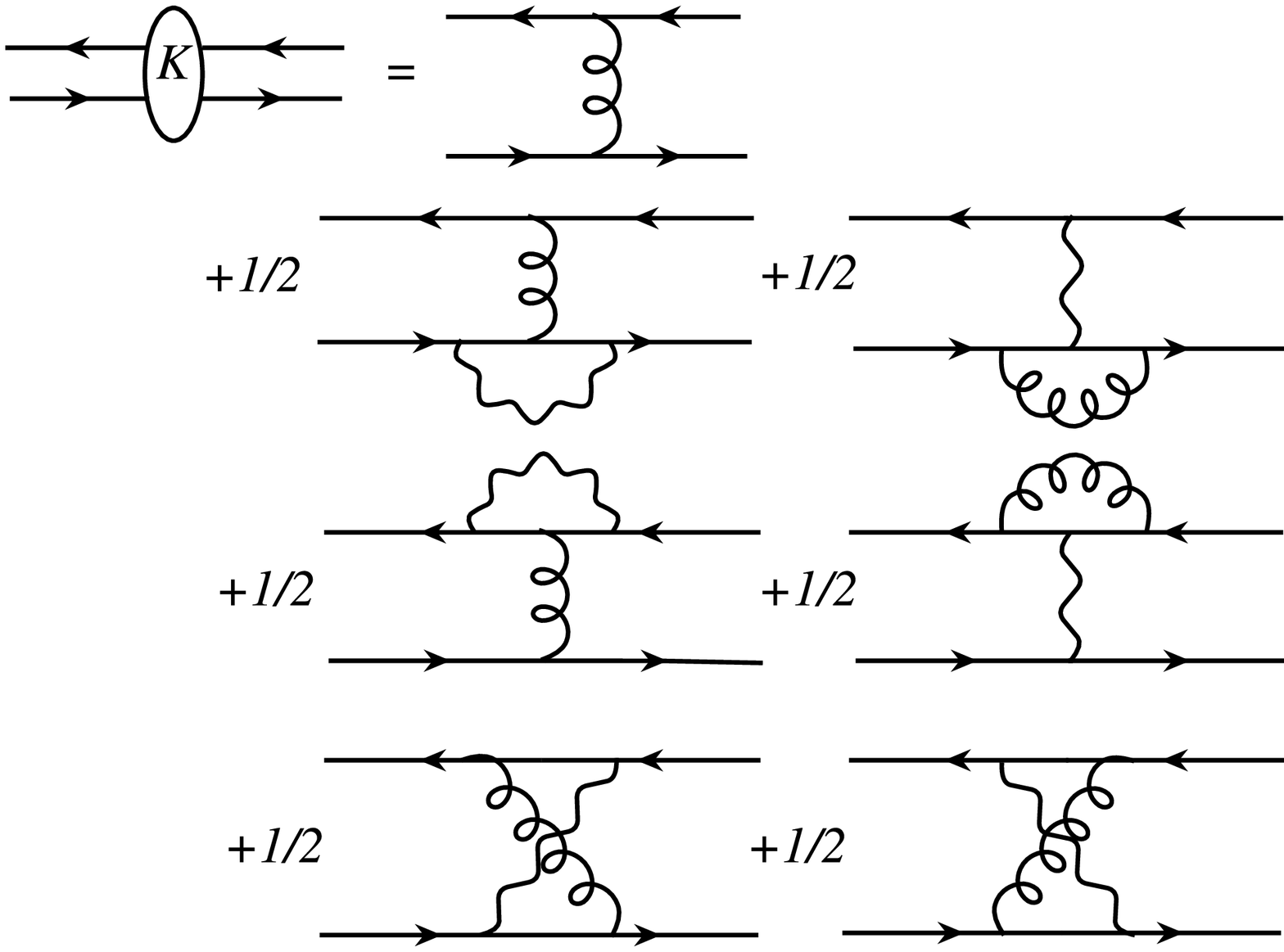,width=10cm}}
\caption{\label{fig:kern}The 2-loop \BS kernel of the hybrid model
that utilizes two different representations of effective gluon
exchange: $\Delta$ denoted by springs, and $\ov{\Delta}$ denoted
by wavy lines.}
\end{figure}

The constructive scheme of \Ref{Bender:1996bb} is an example of this
relation as applied order by order to a Feynman diagram expansion for
$\Sigma(p)$.  An internal quark propagator $S(q)$ is removed and
the momentum flow is adjusted to account for injection of momentum $P$ 
at that point.   With a change in sign 
(related to use of \mbox{$\{\ga_5,\ga_\mu\}=0$} in \Eq{eq:axwi1}), this 
provides one term of the BSE kernel.  The number of such contributions
coming from one self-energy diagram is the number of internal quark
propagators.  Hence the rainbow self-energy generates the ladder BSE kernel;
this is the first term in \Fig{fig:kern}.  A 2-loop self-energy 
diagram (i.e., from 1-loop vertex dressing as in the
present case) generates 3 terms for the BSE kernel.   By substitution into
\Eq{eq:axwi1}  one can confirm that the axial-vector \WT 
identity is preserved.  Similarly, the vector \WT identity is also preserved. 

In the present study, our combined use of two different gluon propagator 
models requires that the BSE kernel obtained so far be subjected to an 
additional procedure to preserve charge conjugation symmetry.    The 
behavior of the \DS equation under charge conjugation
leads to invariance of the self-energy amplitudes under interchange of the two 
different gluon propagators.  In particular, \Eq{eq:qdse} may be written in the
equivalent form
\be
S^{-1}(p)=\imath\slr{p} + m + \int\dk{k}\Delta_{\mu\nu}(p-k)
        \left\{ \frac{4}{3}\,\ga_{\mu}\,S(k)\,\ga_{\nu} + 
        \frac{G}{6}\,\ga_{\ro}\,S(p)\,\ga_{\mu}\,S(k)\,\ga_{\ro}
        \,S(k)\,\ga_{\nu} \right\}~~~,
\label{eq:altqdse}
\ee
and it makes no difference which of the two gluon
propagator models is considered to be implementing the 1-loop vertex dressing.
However, once a quark propagator is removed from a self-energy diagram 
to produce
a contribution to the BSE kernel, interchange of the two
gluon lines, or reversal of the quark line directions, produces a distinct 
contribution to the BSE kernel.  The 
required invariance of the BSE kernel to charge conjugation can be restored 
if the 3 diagrams are symmetrized with respect to interchange of gluon lines,
a procedure that does not alter the quark propagator.  Thus with the present  
hybrid model there are three pairs of 2-loop diagrams to be used for the BSE
kernel; these are displayed in \Fig{fig:kern}.   

With these considerations, the explicit form for the  two-loop BSE kernel 
of the present model is
\bea
\lefteqn{K(p,k;P)_{\al\beta;\de\ga}= -\frac{4}{3}\Delta_{\mu\nu}(p-k)
\left[\ga_{\mu}\right]_{\al\ga}\left[\ga_{\nu}\right]_{\de\beta}}\nonumber\\
&&-\frac{1}{9}\int\dk{q}\left[\Delta_{\mu\nu}(p-k)\ov{\Delta}_{\ro\la}(q)+
\ov{\Delta}_{\mu\nu}(p-k)\Delta_{\ro\la}(q)\right]
\left[\ga_{\mu}\right]_{\al\ga}\left[\ga_{\ro}S(k_-+q)\ga_{\nu}S(p_-+q)
\ga_{\la}\right]_{\de\beta}\nonumber\\
&&-\frac{1}{9}\int\dk{q}\left[\Delta_{\mu\nu}(p-k)\ov{\Delta}_{\ro\la}(q)+
\ov{\Delta}_{\mu\nu}(p-k)\Delta_{\ro\la}(q)\right]
\left[\ga_{\ro}S(p_++q)\ga_{\mu}S(k_++q)\ga_{\la}\right]_{\al\ga}\left[
\ga_{\nu}\right]_{\de\beta}\nonumber\\
&&-\frac{1}{9}\int\dk{q}\left[\Delta_{\ro\la}(q)\ov{\Delta}_{\mu\nu}(k-p-q)+
\ov{\Delta}_{\ro\la}(q)\Delta_{\mu\nu}(k-p-q)\right]
\times\nonumber\\&&\;\;\;\;\;\;\;\;\;\;\;\;\;\;\;\;
\left[\ga_{\ro}S(p_++q)\ga_{\mu}\right]_{\al\ga}\left[\ga_{\la}S(k_--q)
\ga_{\nu}\right]_{\de\beta}~~~.
\label{eq:kern}
\eea
It is straightforward to verify by direct substitution of this kernel and the
1-loop dressed vertex that the axial-vector \WT as expressed by 
\Eq{eq:axwi1} is satisfied.  
The charge conjugation invariance of the kernel can be verified by
substitution into the bound state BSE 
\be
\G(p;P)_{\al\beta}=\int\dk{k}K(p,k;P)_{\al\beta;\de\ga}\;\chi(k;P)_{\ga\de}~~~,
\label{BSE}
\ee
followed by charge conjugation in the form
\be
 C\,\G^T(-p;P)\,C^{-1} = {\cal C}_M \;\G(p;P)~~~. 
\ee
Here $C=-\ga_2\ga_4$ is the charge conjugation matrix, ${\cal C}_M$ is 
the charge parity of the meson,
and $\chi(p;P) = S(p_+)\,\G(p;P)\,S(p_-)$ is the \BS wave function.  

\section{\label{sec:quark}Vertex Correction to the Quark Propagator}

The dressed quark propagator can be represented by a pair of amplitudes and
the two convenient forms used herein are defined by 
\be
S(p)=[\imath\slr{p}A(p^2)+B(p^2)]^{-1}=-\imath\slr{p}\si_V(p^2)+\si_S(p^2)~~~.
\label{eq:qprop}
\ee
Projection of the \DS equation (\ref{eq:qdse}) of the present model 
on to the amplitudes $A$ and $B$ yields the coupled equations
\bea
A(x) &=& I_1(x) + \frac{A(x)\,I_2(x)+B(x)\,I_3(x)}{xA(x)^2+B(x)^2}
\label{eq:q1a}\\
B(x) &=& J_1(x) + \frac{A(x)\,J_2(x)+B(x)\,J_3(x)}{xA(x)^2+B(x)^2}
\label{eq:q1b}
\eea
where \mbox{$x=p^2$}.   In 
Appendix~\ref{sec:app} explicit expressions are given for $I_i(x)$ 
and $J_i(x)$ as integrals involving $\sigma_V(y)$ and $\sigma_S(y)$ for all 
spacelike \mbox{$y=k^2$}.   For \mbox{$i=2,3$}, the integrals $I_i, J_i$
are proportional to the strength $G$ for
vertex dressing.  In the above they multiply scalar and vector propagator 
amplitudes at $x$ because, due to the delta function, one quark propagator
involved in the 2-loop self-energy is always evaluated at the external
momentum.   The quark \DS  equation thus has partly an algebraic structure
and partly an integral structure.

The rainbow truncation, \mbox{$G \to 0$}, yields \mbox{$A(x) =  I_1(x)$} and
\mbox{$B(x) =  J_1(x)$}; the nonlinearity here is contained in the expressions 
for $I_1(x)$ and $J_1(x)$ given  in Appendix~\ref{sec:app}.  In this limit,
our results reduce to those of Ref.\cite{Alkofer:2002bp} thus providing a
useful check of our solutions. 
The limit $\w^2\rightarrow 0$ provides a complementary check: the
distinct effective interactions of the present model, $\Delta_{\mu \nu}(q)$ 
and 
$\ov{\Delta}_{\mu \nu}(q)$, become identical.  In this case the quark DSE,
\Eqs{eq:q1a} and (\ref{eq:q1b}), takes on the algebraic structure that follows
from use of the MN delta-function model in all aspects; this has been applied 
in Ref.~\cite{Bender:1996bb} and in the 1-loop vertex
considerations of Ref.~\cite{Bender:2002as}.

The spacelike ($x>0$) solutions of \Eqs{eq:q1a} and (\ref{eq:q1b}) are 
calculated by iteration subject to the boundary
conditions \mbox{$A(x)\rightarrow 1$} and \mbox{$B(x)\rightarrow m$} for
large enough spacelike $x$.  We note that the solution method can be based
on purely numerical iteration or upon a mixture of numerical iteration 
followed by determination of polynomial roots.  The latter case 
emphasizes the algebraic dependence of \Eqs{eq:q1a} and (\ref{eq:q1b}) upon 
the 
explicit $A(x)$ and $B(x)$ once the integrals $I_i$ and $J_i$ have been 
evaluated.  
Both methods were used as a check.

\begin{figure}[t]
\mbox{\epsfig{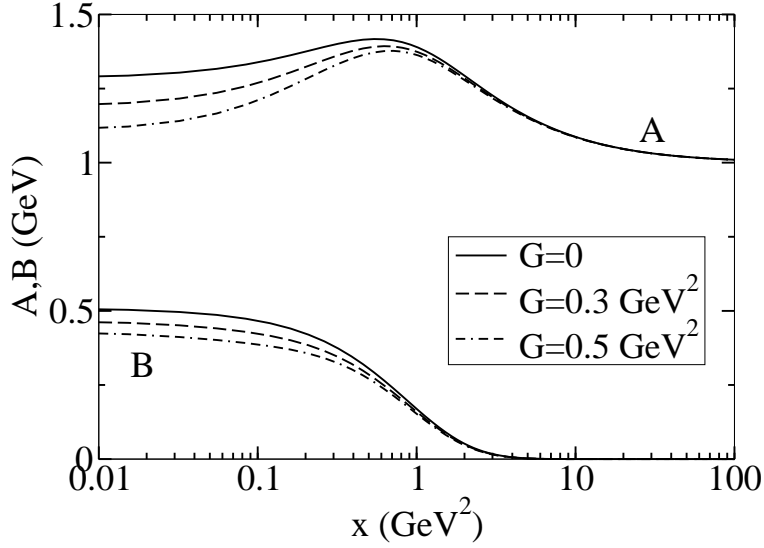}}
\caption{\label{fig:qdses}Spacelike quark propagator functions $A(x)$ 
(top curves) and $B(x)$ (lower curves) in the chiral limit, $m=0$.  
The rainbow truncation (solid line) is compared to the 1-loop dressed 
vertex model with two representative strength parameters.}
\end{figure}

The amplitudes $A,B$ in the spacelike region are shown in \Fig{fig:qdses} for 
the chiral ($m=0$) limit.   The rainbow result is displayed as a solid line
and is compared to results that include vertex dressing characterized by 
strength parameter $G$, with \mbox{$G=0.5~{\rm GeV}^2$} being our estimate 
of the internally consistent value for this model.    The influence
of vertex dressing is evident mainly in the infrared region 
\mbox{$x < 1~{\rm GeV}^2$} where both $A(x)$ and $B$ are 
reduced without an alteration in the qualitative behavior.   There is 
a modest increase in the dynamical mass $B/A$.   These results parallel 
those obtained from the MN delta-function model~\cite{Bender:2002as}.   
At \mbox{$x=0$}, 
$B$ decreases by $\sim$ 20\% and $A - 1$ decreases by $\sim$ 30\%.   

The effect of vertex dressing on the quark propagator can be cast in a more 
physical context by considering the chiral vacuum quark condensate 
$\langle \ov{q}q\rangle^0$ which characterizes dynamical chiral symmetry 
breaking.   In general the condensate at renormalization scale $\mu$ is 
\be
\langle\ov{q}q\rangle^0 = - \lim_{\Lambda\to \infty} Z_4(\mu^2,\Lambda^2)\, 
  N_c\, Tr_D \int^\Lambda \dk{k}S_0(k)~~~,
\label{cond_gen}
\ee
where $S_0(k)$ is the chiral limit propagator.    For an extensive discussion
of the condensate in QCD, see Ref.~\cite{Langfeld:2003ye}. 
The present simple model is
ultraviolet convergent, regularization is not required, and \mbox{$Z_4 =1$}.
\Eq{cond_gen} then yields
\be
\langle\ov{q}q\rangle^0 = 
-\frac{3}{4\pi^2}\int_0^{\infty}dy\,y \frac{B_0(y)}{yA_0(y)^2+B_0(y)^2}~~~,
\label{cond_here}
\ee
which can be considered to be at the typical
hadronic scale \mbox{$\mu \sim 1~{\rm GeV}$}.   The results for 
$\langle\ov{q}q\rangle^0$ given in Table~\ref{tab:qqfpi} show a decrease
of 2\% as the rainbow truncation (\mbox{$G=0.0~{\rm GeV}^2$}) 
is extended by the range of vertex dressing considered here.  

Since the 2-loop BSE kernel used here preserves the axial-vector \WT 
identity, the resulting mass relation~\cite{Maris:1998hd} for 
pseudoscalar mesons is also preserved.   At low $m_q$ this relation
becomes the GMOR relation 
\be
(\hat{f}_{\pi}^0)^2\, m_{\pi}^2=-2\,m_q\langle\ov{q}q\rangle^0+O(m_q^2)~~~,
\label{eq:gmor1}
\ee
where \mbox{$\hat{f}^0_{\pi} = f^0_{\pi}/\sqrt{2}$}, and  $f^0_{\pi}$ 
is the chiral limit leptonic decay constant in the
convention where \mbox{$f_\pi^{\rm expt} = 0.131$}~GeV at 
the physical $u/d$-quark mass.  From previous 
explorations~\cite{Bender:1996bb,Bender:2002as}, we expect $m_\pi$
to be quite stable to vertex dressing;  the relative insensitivity of  
$\langle\ov{q}q\rangle^0$ should therefore be evident in $f^0_{\pi}$ also.   
Here we estimate $f^0_{\pi}$ by 
using the dominant term $i \ga_5 \, B_0(q^2)/f^0_{\pi}$  of 
the chiral pion \BS amplitude~\cite{Roberts:1996hh}
\bea
(f^0_{\pi})^2 &=& 24\int\dk{q}B_0(q^2)\left[\frac{q^2}{2}(\si'_v\si_s-
                                 \si_v\si'_s)+\si_v\si_s\right]\nonumber\\
&=&\frac{3}{2\pi^2}\int_0^{\infty}dy\,y B_0(y)\left[\frac{y}{2}\left(
                      \si'_v\si_s-\si_v\si'_s\right)+\si_v\si_s\right]~~~.
\eea
This expression is known to provide an underestimate by about 10\%.   

The results for $f^0_{\pi}$ are also given in Table~\ref{tab:qqfpi}.   
The variation between the 1-loop (rainbow) kernel 
(\mbox{$G=0.0~{\rm GeV}^2$}) and the estimate for the physical 2-loop kernel 
(\mbox{$G=0.5~{\rm GeV}^2$}) is 1\%.   The expected stability of $m_\pi$
is confirmed from the BSE solution in Sec.~\ref{sec:bse}.  

\begin{table}[th]
\begin{center}
\caption{\label{tab:qqfpi} The chiral quark condensate 
$\langle\ov{q}q\rangle^0$ and the chiral leptonic decay constant 
$f^0_{\pi}$ for a range of vertex dressing strength $G$ (in GeV$^2$). }

\begin{tabular}{|c||c|c|c|c|c|c||c|}\hline
$G$                   & 0.0 &0.1&0.2&0.3&0.4&0.5 & Expt\\ \hline\hline
$(-\langle\ov{q}q\rangle^0)^{1/3}$&0.2511&0.2505&0.2498&0.2492&0.2485&
0.2478& 0.22-0.24~GeV \\ \hline
$f^0_{\pi}$ &0.1190&0.1188&0.1185&0.1183&0.1180&0.1178 & 0.131~GeV \\ \hline
\end{tabular}
\end{center}
\end{table}

  To facilitate our investigation of the BSE with the 2-loop kernel, 
the continuation to \mbox{$P^2 = - M^2$} for total meson momentum will be
implemented under the real axis approximation~\cite{Jarecke:2002xd}, 
defined as the substitution \mbox{$F(x)\rightarrow F(\Re{x})$}
where $F(x)$ stands for the quark propagator amplitudes $A(x)$ and $B(x)$ as 
they appear in the BSE kernel.  The otherwise necessary, model-exact, method
requires knowledge of the complex momentum plane structure of the BSE and
DSE kernels, and has been developed and tested~\cite{Maris:1999nt} only 
for ladder truncation.
Since the present 2-loop kernel is exploratory and future developments are 
expected, we use the approximate method here.  With a ladder kernel of the 
present type, the real axis approximation  produces values for 
$m_\rho$, $m_{K^\star}$ and $m_\phi$ that are within 4\%, 7\% and 
1\% respectively~\cite{Jarecke:2002xd} of the model-exact values; an 
accuracy of about 10\% is anticipated for masses up to about 
2~GeV~\cite{Jarecke:2002xd}.
\begin{figure}[t]
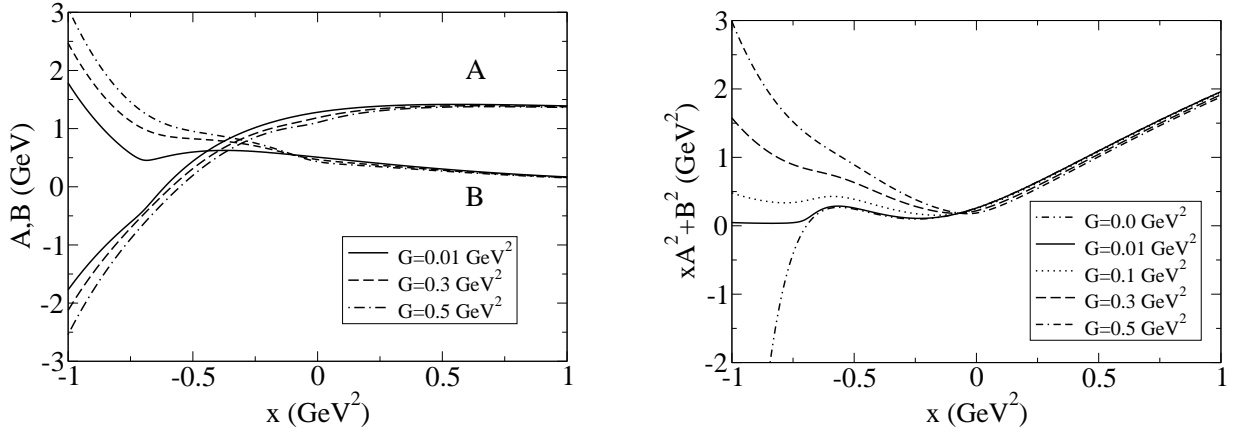

\centering{\ 
\epsfig{figure=realab.eps,width=7.5cm}\ihsp
\epsfig{figure=realden.eps,width=7.5cm} }
\caption{\label{fig:realab} The influence of vertex dressing (\mbox{$G >0$})
on the chiral limit quark propagator for timelike and spacelike real 
\mbox{$x=p^2$}.  {\it Left Panel}: Propagator amplitudes $A(x)$ and $B(x)$. 
The solid line is a good representation of the rainbow truncation.   
{\it Right Panel}: The denominator function.  Finite vertex dressing 
strength  prevents a zero (a physical mass-shell) in this timelike region. }
\end{figure}

Propagator amplitudes $A$ and $B$, in the chiral limit, and along the real 
timelike $p^2$ axis are displayed in \Fig{fig:realab} for representative
values of the
vertex dressing strength $G$.    For equal mass quarks, the meson mass $M$
and the quark \mbox{$x=p^2$} required by the BSE kernel along the real axis
are related by \mbox{$x > -M^2/4$}.  Thus the timelike range shown in 
\Fig{fig:realab} is sufficient for a meson mass up to 2~GeV.  
The repulsive effect of our model vertex dressing is evident in the 
enhanced dynamical chiral symmetry breaking ($B > 0$) in the timelike 
region.   In models of this type,
the strong dynamical chiral symmetry breaking creates a significant infrared
timelike domain where the dressed quark has no real mass shell;  $\bar q q$
bound states with masses within this domain have no spurious $\bar q q$
decay mode.    We illustrate this in
\Fig{fig:realab} which displays the propagator denominator $xA(x)^2 + B(x)^2$
in the chiral limit.  A zero of this quantity indicates a quark mass shell.  
In rainbow truncation (\mbox{$G = 0.0$}), the physical domain
of applicability identified this way is 
\mbox{$ x > -0.7~{\rm GeV}^2$}~\cite{Alkofer:2002bp}, which allows meson 
states below about 1.7~GeV to be free of spurious $\bar q q$ widths.  
In \Fig{fig:realab} our model vertex dressing can be seen to enlarge the 
physical domain of applicability; any meson states up to 2~GeV would
be without such spurious widths.   This is consistent with the effective
increase in infrared strength that is generated by vertex dressing.

\section{\label{sec:bse}The \BS Equation and Mesons}

With the constructed 2-loop kernel, \Eq{eq:kern}, the homogeneous \BS equation 
for mesons becomes
\bea
\lefteqn{\G(p;P)=\int\dk{k}K(p,k;P)_{\al\beta;\de\ga}\chi(k;P)_{\ga\de}}\nonumber\\
&=&-\frac{4}{3}\int\dk{k}\Delta_{\mu\nu}(p-k)\ga_{\mu}\chi(k;P)\ga_{\nu}\nonumber\\
&&-\frac{G}{12}\int\dk{k}\Delta_{\mu\nu}(p-k)\left\{\ga_{\mu}\chi(k;P)
\ga_{\ro}S(k_-)\ga_{\nu}S(p_-)\ga_{\ro}+\ga_{\ro}S(p_+)\ga_{\mu}S(k_+)\ga_{\ro}
\chi(k;P)\ga_{\nu}\right\}\nonumber\\
&&-\frac{G}{12}\int\dk{k}\Delta_{\mu\nu}(p-k)\left\{\ga_{\mu}S(k_+)\ga_{\ro}\chi(k;P)
\ga_{\nu}S(p_-)\ga_{\ro}+\ga_{\ro}S(p_+)\ga_{\mu}\chi(k;P)\ga_{\ro}S(k_-)\ga_{\nu}
\right\}\nonumber\\
&&-\frac{G}{12}\int\dk{k}\Delta_{\mu\nu}(p-k)\left\{\ga_{\ro}\chi(p;P)
\ga_{\mu}S(k_-)\ga_{\ro}S(k_-)\ga_{\nu}+\ga_{\mu}S(k_+)\ga_{\ro}S(k_+)\ga_{\nu}
\chi(p;P)\ga_{\ro}\right\}\nonumber\\
\label{eq:bse1}
\eea
where the integrations over the $\de$-functions have been carried out and 
$\chi(p;P)=S(p_+)\G(p;P)S(p_-)$.  Discrete solutions for meson mass $M_n$ 
exist for $P^2=-M_n^2$.

To solve for a particular meson, one must specify the appropriate quantum
numbers $J^{P C}$ by expressing the bound state vertex $\Gamma$ in the 
general covariant form that has the corresponding transformation properties. 
The quark flavor content can be specified by the current quark masses.   Here
we consider pseudoscalar, vector and axial-vector charge eigenstates and  
adopt isospin symmetry through degenerate $u$ and $d$ quarks. 
The general covariant forms used here are
\bea
\G^{PS}(p;P)&=&\ga_{5}\left[\G_0(p;P)-\imath\slr{P}\G_1(p;P)-\imath\slr{p}
\G_2(p;P)-\left[\slr{P},\slr{p}\right]\G_3(p;P)\right]~~,
\label{eq:pseu}\\
\G_{\mu}^{V}(p;P) &=& \ga_{\mu}^T \left[
\imath\G_0(p;P)+\slr{P}\G_{1}(p;P)-\slr{p}\G_2(p;P)+\imath\left[\slr{P},
\slr{p}\right]\G_3(p;P)\right]\nonumber\\
&&+ p_\mu^T \left[\G_{2}(p;P)+2\imath
\slr{P}\G_3(p;P) \right]\nonumber\\
&&+ p_\mu^T \left[\G_4(p;P)+
\imath\slr{P}\G_5(p;P)-\imath\slr{p}\G_6(p;P)+\left[\slr{P},\slr{p}\right]
\G_7(p;P)\right]~~,
\label{eq:vect}\\
\G_{\mu}^{AV}(p;P) &=& \ga_5 \ga_{\mu}^T
\left[\imath\G_0(p;P)+\slr{P}\G_{1}(p;P)-\slr{p}\G_2(p;P)+\imath\left[\slr{P},
\slr{p}\right]\G_3(p;P)\right]\nonumber\\
&&+\ga_5 p_\mu^T \left[\G_{2}(p;P)+2
\imath\slr{P}\G_3(p;P) \right]\nonumber\\
&&+\ga_5 p_\mu^T \left[\G_4(p;P)+
\imath\slr{P}\G_5(p;P)-\imath\slr{p}\G_6(p;P)+\left[\slr{P},\slr{p}
\right]\G_7(p;P)\right]~~,
\label{eq:axvect}
\eea
where the notation $a_\mu^T$ denotes a 4-vector transverse to $P_\mu$,
i.e., \mbox{$a_\mu^T =$} \mbox{$ t_{\mu \nu}(P) a_\nu$}. 
The vector and axial-vector amplitudes $\Gamma_\mu^{V/AV}$ are transverse 
to the 
total momentum $P$.  The $\G_i$ are scalar functions of $p^2$, $p\s P$.  
For charge eigenstates, the $\G_i$ are either odd or even 
under the interchange $p\s P\rightarrow-p\s P$.

For numerical solution of \Eq{eq:bse1} we project onto the basis of Dirac 
covariants to obtain coupled equations for the amplitudes $\G_i$.  The 
dimensionality of integration is reduced by expansion of $\G_i$
in the complete orthonormal set of Chebyshev polynomials  $T_k(z)$ in 
the variable $z=p\s P/\sqrt{p^2P^2}$,
\be
\G_i(p^2,z;P^2)=\sum_{k=0}^{N_{ch}}\, i^k\, T_k(z)\, \G_i^k(p^2;P^2)~~~.
\label{eq:chebyexp}
\ee
Projection of the equation onto the $T_k$ basis then produces a set of 
coupled 1-dimensional integral equations for the  $\G_i^k(p^2;P^2)$.
An explicit factor of $p\s P$ is extracted from the odd $\G_i$ and
an even order Chebyshev expansion is used for the remaining factor.  
Convergence and stability is normally achieved with $N_{ch}=4$ or $6$.  
The meson mass $M_n$ is determined by introducing a linear eigenvalue 
$\lambda(M)$ so that the \BS equation reads 
\be
\lambda(M)\G(p;P)_{\al\beta} = \int\dk{k}\, K_{\al\beta;\de\ga}(p,k;P)\, 
                           \left[S(k_-)\G(k;P)S(k_+)\right]_{\ga\de}~~~, 
\ee
with $P^2=-M^2$; the mass $M$ is varied until $\la(M=M_n)=1$.   The physical 
ground state in a given channel is determined by the largest real eigenvalue.

\section{\label{sec:res}Numerical Results and Discussion}

We explore ground state $u/d$-quark mesons and adopt the ladder kernel 
model of \Ref{Alkofer:2002bp} which provides the
parameters $\w=0.5~{\rm GeV}, D=16~{\rm GeV}^{-2}$ (see \Eq{eq:FRgluon}) 
and $m_{u/d}=5$~MeV.
The strength parameter $G$ for vertex dressing is varied in the range 
0.0-0.5~GeV$^2$.   Our model \BS kernel produces real ground state masses
since it does not contain a mechanism for a hadronic decay width; it also
does not distinguish between isoscalar and isovector states.   We 
indicate the physical state identifications with these qualifications
in mind. 
Real mass solutions of the type obtained here have in the past provided
a basis for successful perturbative descriptions of strong decay widths 
in non-scalar channels, an example being the \mbox{$\rho \rightarrow \pi \pi$} 
width~\cite{Praschifka:1987pt,Hollenberg:1992nj,Mitchell:1997dn}.

\begin{table}[th]
\begin{center}
\caption{\label{tab:ps+vect} The \mbox{$0^{-+}$} ($\pi$) and \mbox{$1^{--}$} 
($\rho$) masses in GeV. }

\begin{tabular}{|cc|cccccc|c|}\hline
\multicolumn{2}{|c|}{$G$~(GeV$^2$)} &0&0.1&0.2&0.3&0.4&0.5& Expt\\\hline\hline
$N_{\rm ch}=$ 2 \rule{2mm}{0ex} & $m_\pi$ &\rule{3mm}{0ex}0.143\rule{3mm}{0ex} 
                                          &\rule{3mm}{0ex}0.142\rule{3mm}{0ex}
                                          &\rule{3mm}{0ex}0.142\rule{3mm}{0ex}
                                          &\rule{3mm}{0ex}0.141\rule{3mm}{0ex}
                                          &\rule{3mm}{0ex}0.141\rule{3mm}{0ex}
                                          &\rule{3mm}{0ex}0.140\rule{3mm}{0ex}
                                          & \\
                & $m_\rho$& 0.806 &0.808 &0.810 &0.811 &0.812 &0.813&\\\hline
$N_{\rm ch}=$ 4 \rule{2mm}{0ex} & $m_\pi$ & 0.140&0.140&0.140&0.140&0.140&
0.139&\\
                & $m_\rho$& 0.794&0.791&0.788&0.783&0.777&0.771&\\\hline
$N_{\rm ch}=$ 6 \rule{2mm}{0ex} & $m_\pi$ & 0.140&0.140&0.140&0.140&0.141&
0.139& \rule{3mm}{0ex}0.139\rule{3mm}{0ex}\\
                & $m_\rho$& 0.794&0.790&0.785&0.779& 0.772&0.763&
        \rule{3mm}{0ex}0.770\rule{3mm}{0ex}\\
\hline\hline
\end{tabular}
\end{center}
\end{table}

Table~\ref{tab:ps+vect} records the pion (\mbox{$0^{-+}$}) mass values for 
vertex 
dressing strengths between $G=0.0~{\rm GeV}^2$ (rainbow-ladder truncation) 
and \mbox{$G = 0.5~{\rm GeV}^2$} (our estimate of the physical value).
The mass is practically insensitive to model details; this illustration
of chiral symmetry preservation through the axial-vector \WT identity
has been observed in related earlier work with a simplified ladder 
kernel~\cite{Bender:1996bb,Bender:2002as,Bhagwat:2004hn}.  
Table~\ref{tab:ps+vect} also 
illustrates the stability and convergence with respect to $N_{\rm ch}$ (the 
number of Chebyschev polynomials used to expand the angle dependence of 
the \BS amplitudes according to \Eq{eq:chebyexp}).  This and other 
numerical techniques were applied until an accuracy of better than 5\%
was achieved for the masses.
\begin{figure}[th]
\mbox{\epsfig{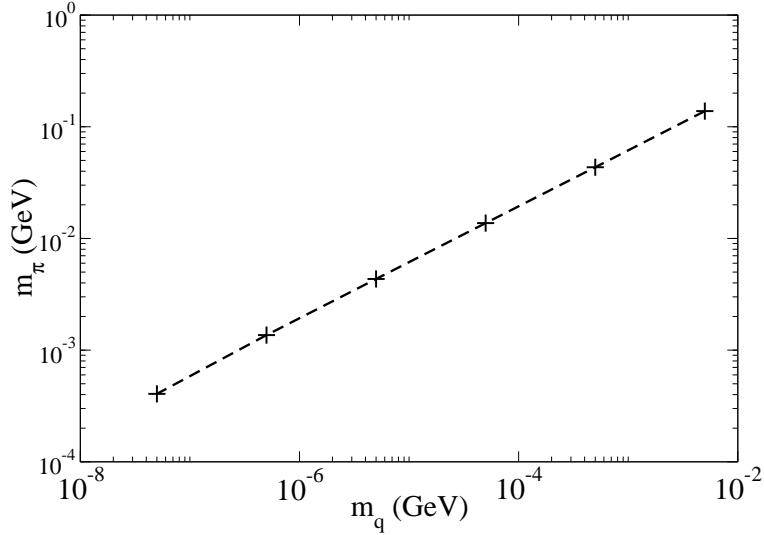}}
\caption{\label{fig:chiralpi} The $m_q$ dependence of the pion mass near
the chiral limit.   The fitted dashed line confirms that the dressed 
vertex model (characterized by $G=0.5~{\rm GeV}^2$) satisfies the GMOR 
relation, \protect\Eq{eq:gmor1}. }
\end{figure}

With strength $G=0.5~{\rm GeV}^2$ for vertex dressing, the 
$m_q$-dependence of the pseudoscalar bound state mass near the chiral limit
is displayed in \Fig{fig:chiralpi}.   The form
\be
m_\pi=a_1\sqrt{m_q}+a_2 \, m_q
\ee
provides a good fit with $a_1=1.94~{\rm GeV}^{\frac{1}{2}}$, and $a_2=0.231$.  
The dominance of the first term (by a factor of about 10$^3$) suggests 
that the GMOR mass relation,
\Eq{eq:gmor1}, has been reproduced.    This is reinforced by noting that,
according to \Eq{eq:gmor1}, we should have the correspondance  
\mbox{$a_1 \sim 2\sqrt{|\langle\ov{q}q\rangle^0|}/f^0_\pi$}  
and with the model value of $\langle\ov{q}q\rangle^0$ presented in 
Table~\ref{tab:qqfpi}, we deduce from the fitted $a_1$ that 
\mbox{$f^0_{\pi}=0.1272$}~GeV.  
This chiral value is 3\% lower than the experimental value 
$f_{\pi}=0.1307$~GeV (at the physical mass) and is consistent with 
previous findings~\cite{Maris:1997tm}.   It is also consistent with 
the independent calculation of $f^0_{\pi}$ in Section~\ref{sec:quark}. 
\begin{figure}[th]
\mbox{\epsfig{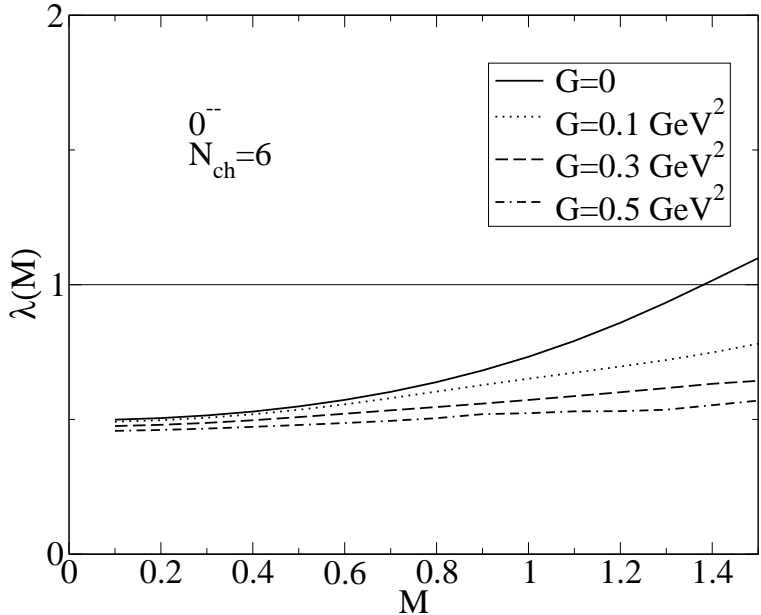}}
\caption{\label{fig:psm} Exotic pseudoscalar ($0^{--}$) meson \BS
eigenvalue from the ladder kernel ($G=0$) and from the dressed vertex model
($G > 0$).  }
\end{figure}

\Fig{fig:psm}
displays the eigenvalue for the exotic $0^{--}$ channel where the ladder
kernel produces a mass of 1.38~GeV.  Our model vertex dressing has a
large repulsive effect; the physical value for $G=0.5$~GeV suggests that 
any mass solution would be above 2~GeV. 
There is a scalar $0^{++}$ solution to the present model at a mass of 0.7~GeV
in ladder approximation; this decreases to 0.6~GeV for the physical 2-loop 
kernel.  The ladder value is consistent with previous models of this 
type~\cite{Maris:2000ig,Burden:2002ps,Alkofer:2002bp}.  
A physical identification for the $0^{++}$ solution here is not appropriate 
because hadronic configurations such as $\bar K K$ are expected to play
an important role, and the ladder-rainbow  truncation is known to be 
deficient in this channel~\cite{Bender:1996bb}. 
\begin{figure}[h]
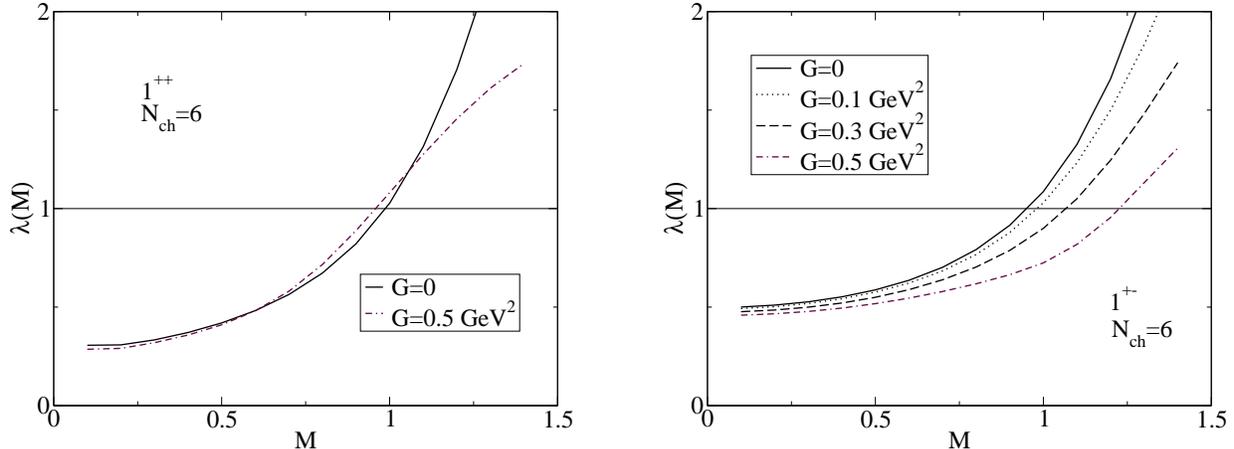

\centering{\
\epsfig{figure=finaxp.eps,width=7.5cm}\ihsp
\epsfig{figure=axmn6.eps,width=7.5cm} }
\caption{\label{fig:axp}  Axialvector  meson \BS eigenvalue   
from the  ladder kernel (\mbox{$G=0$}) and from the dressed vertex model 
(\mbox{$G > 0$}).  {\it Left Panel}: The $1^{++}$ ($a_1/f_1$)
state.  {\it Right Panel}: The $1^{+-}$ ($b_1/h_1$) state. }
\end{figure}
 
The mass values for the \mbox{$1^{--}$} ($\rho/\omega$) channel 
are presented in Table~\ref{tab:ps+vect} for the relevant
range of parameters $G$ and $N_{ch}$.     The effect of the present
model of 1-loop
vertex dressing is seen to be an attraction of only 30~MeV.   In contrast,
a repulsive change of about 70~MeV applies when the delta function 
representation
is employed for all aspects of the 2-loop kernel~\cite{Bender:2002as}.
That simplification also provided an extension to the complete ladder 
summation of the one-gluon exchange mechanism for the vertex; the result 
being just a 75~MeV repulsion~\cite{Bender:2002as}.   Evidently, 
finite momentum range effects in the kernel can influence the net result
from vertex dressing and whether there is net attraction or repulsion. 

However, the larger issue in the present context is why the typical 
effect beyond
rainbow-ladder seems to be no more than $\pm$10\% for the ground state 
vector mass when there is no explicit protection from an underlying
symmetry.  With the pseudoscalar state fixed by chiral symmetry, the 
hyperfine splitting that can be generated by modeling  the quark-gluon
vertex by effective gluon exchange is evidently a weak perturbation on
the ladder kernel.  Recent work indicates that significant attraction
is provided to the dressed gluon-quark vertex by the triple-gluon vertex,
and a schematic model implementation suggests that up to  a 30\%
attractive effect on $m_\rho$ can be provided in this way beyond 
rainbow-ladder~\cite{Bhagwat:2004hn}.   Further work is required
for an understanding of  how best to model the quark-gluon vertex for 
hadron physics.  

No mass solution below 2~GeV was found in the exotic channel 
$1^{-+}$.   In this mass range, the largest eigenvalue had reached ~0.5-0.8.
Masses above 2~GeV might be possible in a model of this type.
\begin{table}[th]
\begin{center}
\caption{\label{tab:axvects} The \mbox{$1^{++}$} ($a_1/f_1$) and 
\mbox{$1^{+-}$} ($b_1/h_1$) masses in GeV. }

\begin{tabular}{|cc|cccccc|c|}\hline
\multicolumn{2}{|c|}{$G$~(GeV$^2$)} &0&0.1&0.2&0.3&0.4&0.5& Expt\\\hline\hline
$N_{\rm ch}=$ 2 \rule{2mm}{0ex} & $m_{a_1}$ &\rule{3mm}{0ex}1.061\rule{3mm}{0ex} 
                                          &\rule{3mm}{0ex}1.062\rule{3mm}{0ex}
                                          &\rule{3mm}{0ex}1.060\rule{3mm}{0ex}
                                          &\rule{3mm}{0ex}1.056\rule{3mm}{0ex}
                                          &\rule{3mm}{0ex}1.050\rule{3mm}{0ex}
                                          &\rule{3mm}{0ex}1.043\rule{3mm}{0ex}
                                          & \\
                & $m_{b_1}$& 0.968 &0.983 &0.999 &1.017 &1.038 &1.064&\\\hline
$N_{\rm ch}=$ 4 \rule{2mm}{0ex} & $m_{a_1}$& 0.989&0.985&0.980&0.975&0.969&
0.964&\\
                & $m_{b_1}$& 0.952&0.978&1.008&1.046&1.094&1.155&\\\hline
$N_{\rm ch}=$ 6 \rule{2mm}{0ex} & $m_{a_1}$& 0.989&0.982&0.975&0.969&0.963&
0.959&\rule{3mm}{0ex}1.230\rule{3mm}{0ex}\\
                & $m_{b_1}$& 0.954&0.983&1.020&1.070&1.138&1.227&\\\hline
$N_{\rm ch}=$ 8 \rule{2mm}{0ex} & $m_{b_1}$& 0.954&0.984&1.024&1.080&1.161&
1.243&\rule{3mm}{0ex}1.230\rule{3mm}{0ex}\\
\hline\hline
\end{tabular}
\end{center}
\end{table}

The eigenvalue behavior for the axial vector solutions in  the 
\mbox{$1^{++}$} ($a_1/f_1$) and 
\mbox{$1^{+-}$} ($b_1/h_1$) channels are displayed in \Fig{fig:axp} 
and also in Table~\ref{tab:axvects}.   In previous work, 
the ladder truncation, constrained by 
chiral data, is generally found to be 200-400~MeV too attractive for 
these P-wave states~\cite{MarisPrivCom,Jarecke:2002xd,Alkofer:2002bp}. 
Our present results agree with this.
The \mbox{$1^{++}$} channel shows a 30~MeV of attraction
due to the effect of 1-loop dressing added to the ladder kernel.   However,
in the \mbox{$1^{+-}$} ($b_1/h_1$) channel, we find a repulsive effect
of 290~MeV above the ladder kernel result, yielding a value close
to $m^{\rm expt}_{{\rm b}_1}$.   We are not able to compare 
these findings with previous work on vertex dressing since such P-wave 
states do not have solutions in the models considered previously for
that purpose~\cite{Bender:1996bb,Bender:2002as,Bhagwat:2004hn}.
Other studies of $a_1$ and $b_1$ based on the ladder-rainbow truncation
have used a separable approximation where the quark propagators are the 
phenomenological instruments~\cite{Burden:1997nh,Bloch:1999vk,Burden:2002ps}, 
these studies find more acceptable masses for both states in the vicinity of 
1.3~GeV.

The typical widths of the ground state axial vector mesons are
about 20\% of the mass and 3$\pi$ and $4\pi$ decay channels are prominent.
Although widths of this magnitude  have been successfully generated 
perturbatively for vectors~\cite{Jarecke:2002xd} and
axial-vectors~\cite{Bloch:1999vk} from BSE solutions that do not have 
the decay channels within the kernel, it is possible that such dynamics 
are responsible for important contributions to the masses beyond 
ladder-rainbow truncation.

\section{\label{sec:conc}Summary}

The effect of a 1-loop model of quark-gluon vertex dressing on the masses of 
the light quark  pseudoscalar, vector and axial-vector mesons has been 
studied.  To facilitate calculations the vertex model consists
of an effective single gluon exchange represented by a momentum 
$\de$-function with strength $G$.  The ladder-rainbow term is represented 
by a finite range effective gluon exchange.  
The axial-vector \WT identity and charge conjugation 
properties are used to construct a consistent \BS kernel from the quark 
self-energy.   This model extends recent explorations beyond the 
rainbow-ladder level by  allowing axial-vector meson solutions.

At spacelike momenta the vertex 
dressing results in only slight corrections to the quark propagator.  
The \BS equation was solved numerically for the ground state pseudoscalar, 
vector and axial-vector mesons.  The results show that
corrections to the rainbow-ladder truncation generate very little 
change to pseudoscalar masses (and effectively vector masses) when they
are defined consistently with chiral symmetry.  
  
The axial-vector mesons respond to the present vertex dressing model in
a way that calls for further study.    
The mass of the $1^{+-}$ ($b_1/h_1$) state, too small by about 300~MeV 
in rainbow-ladder truncation,  was raised by 290~MeV by the vertex dressing
thus giving a satisfactory value.   However the \mbox{$1^{++}$} ($a_1/f_1$) 
state decreased in mass by about 30~MeV, leaving it still about 300~MeV 
too light compared to experiment.  In the exotic meson channels, $0^{--}$ 
and $1^{-+}$, the vertex dressing 
effects are repulsive; possible solutions are well above 1.5~GeV and
beyond the range of the methods of the present study.

There is very little information on the non-perturbative structure of
the dressed quark-gluon vertex that can be used to guide a practical
phenomenology.  The Ball-Chiu (Abelian) Ansatz~\cite{Ball:1980ay}, 
often-used to generate
the quark self-energy, is not useful here because the only known way to 
define a chiral-symmetry-preserving BSE kernel requires
an explicit Feynman diagram representation of the self-energy.  The present
model adopts a one gluon exchange vertex structure motivated by a previous
study.  Dynamical information on the non-perturbative structure of the 
triple gluon vertex, and the contribution it can make to the quark-gluon 
vertex, is needed to further clarify how best to model the BSE kernel 
beyond ladder-rainbow.  The attractive
influence~\cite{Bhagwat:2004hn} of the triple gluon vertex can have 
important consequences for modeling of the hadron spectrum.  
The high mass meson states are furthermore exptected to move significantly 
when including their hadronic decay channels.

\appendix
\section{\label{sec:app} Details of the Quark \DS Equation}

The quark \DS equation of the present 2-loop model, \Eq{eq:qdse}, 
is
\be
S^{-1}(p)=\imath\slr{p}+m+\int\dk{k}\,\Delta_{\mu\nu}(p-k)\left\{\frac{4}{3}
\gamma_{\mu}S(k)\gamma_{\nu}+\frac{G}{6}\gamma_{\mu}S(k)\gamma_{\ro}S(k)
\gamma_{\nu}S(p)\gamma_{\ro}\right\}~~~.
\label{eq:adse}
\ee
Projection of \Eq{eq:adse} onto the basis of Dirac vector and scalar matrices
produces coupled non-linear equations for the propagator amplitudes
$A$ and $B$ introduced in \Eq{eq:qprop}.  With Lorentz invariants $x=p^2$, 
$y=k^2$, $z=k\s p/\sqrt{k^2p^2}$,  and integration measure 
\mbox{$\int\dk{k}=$} \mbox{$1/(2\pi)^3\int_0^{\infty}dy\,y$} 
\mbox{$\int_{-1}^1dz\,\sqrt{1-z^2}$}, the equation then takes the form,
given in Eqs.~(\ref{eq:q1a}) and (\ref{eq:q1b}),
\bea
A(x) &=& I_1(x) + \frac{A(x)\,I_2(x)+B(x)\,I_3(x)}{xA(x)^2+B(x)^2}
\label{eq:App_A}\\
B(x) &=& J_1(x) + \frac{A(x)\,J_2(x)+B(x)\,J_3(x)}{xA(x)^2+B(x)^2}~~~,
\label{eq:App_B}
\eea
where the $I_i(x),J_i(x)$ are scalar integrals over non-linear combinations
of $A(y)$ and $B(y)$ for spacelike $y$.  Because of the gaussian form 
\Eq{eq:FRgluon} for the kernel $\Delta_{\mu\nu}(p-k)$ in the present model, 
the results of the 
angular integrations required for evaluation of $I_i(x)$ and $J_i(x)$ can be 
expressed in closed form.  To this end one needs the two 
confluent hypergeometric functions~\cite{Abramowitz:1968bk},
\bea
Z_{\mu}^0&=\int_{-1}^1dt\,\exp{\{\mu (t-1)\}}\left(1-t^2\right)^{-\half}&=
\pi {}_1F_1(1/2,1;-2\mu)~~~,\\
Z_{\mu}^1&=\int_{-1}^1dt\,\exp{\{\mu (t-1)\}}\left(1-t^2\right)^{\half}&=
\frac{\pi}{2} {}_1F_1(3/2,3;-2\mu)~~~,
\eea
where \mbox{$\mu=2\sqrt{xy}/\w^2$}.
These hypergeometric functions can be numerically evaluated for general 
complex argument $\mu$ using their series expansion ($|\mu|<1$), direct 
numerical integration ($1<|\mu|<17$), or asymptotic formulae 
($|\mu|>17$)~\cite{Abramowitz:1968bk}.  
There are two important cases: real $\mu$  and purely imaginary $\mu$, 
whereupon the angular integrals reduce to modified Bessel or Bessel 
functions, respectively.  For real $\mu$ (i.e., real \mbox{$x=p^2 > 0$})
\be
Z_{\mu}^0 = \pi e^{-\mu}I_0(\mu)~~~;~~~~~Z_{\mu}^1 = 
                           \frac{\pi}{\mu} e^{-\mu}I_1(\mu)~~~,
\ee
whereas for purely imaginary $\mu$  (i.e., real \mbox{$x=p^2 < 0$})
\be
Z_{\mu}^0 = \pi e^{-\mu}J_0(\mu')~~~;~~~~~Z_{\mu}^1 =
                            \frac{\pi}{\mu'} e^{-\mu}J_1(\mu')~~~,
\ee
where \mbox{$\mu=\imath\mu'$}. The quantities $I_i(x),J_i(x)$ can then
be expressed as the one-dimensional integrals
\bea
I_1(x)&=&1+\frac{2D}{\pi\w^2}\int_0^{\infty}dy\,y\exp{\left(
-\frac{(\sqrt{x}-\sqrt{y})^2}{\w^2}\right)}\si_v(y)\times\nonumber\\&&
\left[Z_{\mu}^0\half\w^2\left(1+\frac{y+2\w^2}{x}\right)+Z_{\mu}^1\left(
-y\left(2+\frac{\w^2}{x}\right)-\w^2\left(1+2\frac{\w^2}{x}\right)\right)
\right]~~~,\nonumber\\
I_2(x)&=&G\frac{2D}{\pi\w^2}\int_0^{\infty}dy\,y\exp{\left(-\frac{
(\sqrt{x}-\sqrt{y})^2}{\w^2}\right)}\times\nonumber\\&&
\!\!\!\!\!\!\!\!\!\!\!\!\!\!\!\!\!\!\!\!\!\!\!\!\!\!\!\!\!\!\!\!
\left\{\si_v^2(y)\left[Z_{\mu}^0\frac{1}{4}\left(y\w^2+\w^4+\frac{y\w^4+
4\w^6}{x}\right)+Z_{\mu}^1\frac{1}{4}\left(-xy-6y\w^2-y^2-2\w^4-\frac{2y\w^4+
4\w^6}{x}\right)\right]\right.\nonumber\\&&
\left.+\si_s^2(y)\left[Z_{\mu}^0\left(-\frac{\w^4}{8x}\right)+
Z_{\mu}^1\frac{\w^4}{4x}\right]\right\}~~~,\nonumber\\
I_3(x)&=&G\frac{2D}{\pi\w^2}\int_0^{\infty}dy\,y\exp{\left(-\frac{
(\sqrt{x}-\sqrt{y})^2}{\w^2}\right)}\si_v(y)\si_s(y)\times\nonumber\\&&
\left[Z_{\mu}^0\frac{1}{8}\w^2\left(-1-\frac{y+3\w^2}{x}\right)+
Z_{\mu}^1\frac{1}{4}\left(2y+\w^2+\frac{y\w^2+3\w^4}{x}\right)\right]~~~,
\nonumber\\
J_1(x)&=&m+\frac{2D}{\pi\w^2}\int_0^{\infty}dy\,y\exp{\left(
-\frac{(\sqrt{x}-\sqrt{y})^2}{\w^2}\right)}\si_s(y)\times\nonumber\\&&
\left[Z_{\mu}^0(-\w^2)+Z_{\mu}^1(x+y+2\w^2)\right]~~~,\nonumber\\
J_2(x)&=&G\frac{2D}{\pi\w^2}\int_0^{\infty}dy\,y\exp{\left(-\frac{
(\sqrt{x}-\sqrt{y})^2}{\w^2}\right)}\si_v(y)\si_s(y)\times\nonumber\\&&
\left[Z_{\mu}^0\frac{1}{8}\left(-x\w^2-y\w^2-3\w^4\right)+Z_{\mu}^1
\frac{1}{4}\left(2xy+x\w^2+y\w^2+3\w^4\right)\right]~~~,\nonumber\\
J_3(x)&=&G\frac{2D}{\pi\w^2}\int_0^{\infty}dy\,y\exp{\left(-\frac{
(\sqrt{x}-\sqrt{y})^2}{\w^2}\right)}\times\nonumber\\&&
\!\!\!\!\!\!\!\!\!\!\!\!\!\!\!\!\!\!\!\!\!\!\!\!\!\!\!\!\!\!\!\!
\left\{\si_v^2(y)\left[Z_{\mu}^0\left(-\frac{\w^4}{8}\right)+Z_{\mu}^1
\frac{\w^4}{4}\right]+
\si_s^2(y)\left[Z_{\mu}^0\frac{\w^2}{4}+Z_{\mu}^1\frac{1}{4}(-x-y-2\w^2)
\right]\right\}~~~.\nonumber\\
\label{eq:ij}
\eea

The equations can be solved numerically by iteration subject to the boundary
conditions \mbox{$A(x)\rightarrow 1$} and \mbox{$B(x)\rightarrow m$} for
large spacelike $x$.  Alternatively, once the spacelike integrals
for $I_i(x),J_i(x)$ have been obtained, one may seek to utilize the
polynomial structure of the equations in the explicitly appearing amplitudes 
$A(x)$ and $B(x)$.   For completeness, we outline the derivation of
the polynomial form.

Elimination of  either $A$ or $B$ from the 
right-hand side of \Eqs{eq:App_A} and (\ref{eq:App_B}) produces
\bea
\left[xA^2+B^2\right]\left[(A-I_1)J_3-(B-J_1)I_3\right]-(I_2J_3-J_2I_3)A&=&0
\label{eq:q2a}\\
\left[xA^2+B^2\right]\left[(B-J_1)I_2-(A-I_1)J_2\right]-(I_2J_3-J_2I_3)B&=&0~~,
\label{eq:q2b}
\eea
where we have dropped the argument $x$ from $A$ and $B$.
This can also be written as
\be
\frac{(I_2J_3-J_2I_3)}{\left[xA^2+B^2\right]}=
\frac{(A-I_1)J_3-(B-J_1)I_3}{A}=\frac{(B-J_1)I_2-(A-I_1)J_2}{B}~~.
\label{eq:q3a}
\ee
The second equality can be multiplied out to give an expression for $B^2$:
\be
B^2I_3=B\left[A(J_3-I_2)+J_1I_3-I_1J_3\right]+A^2J_2+A(J_1I_2-I_1J_2)~~.
\label{eq:q3b}
\ee
This expression is now used to eliminate any factors $B^2$ in 
\Eqs{eq:q2a} and (\ref{eq:q2b}), yielding
\begin{eqnarray*}
B\left[A(E-C)-I_1E\right]+A^2D+A(J_1C-I_1D)-I_3E&=&0~~,\\
B\left[A^2(J_3E+I_2(C-2E))+AE(2I_1I_2-2I_1J_3+I_3J_1)+
E(I_1^2J_3-I_1I_3J_1-I_3^2)\right]&&\\
+A^3J_2(E-C)+A^2(C(I_1J_2-J_1I_2)+E(J_1I_2-2I_1J_2))+
AEI_1(I_1J_2-J_1I_2)&=&0~~,
\end{eqnarray*}
where $C=xI_3^2+I_2^2$, $D=xI_3J_3+I_2J_2$ and $E=I_2J_3-J_2I_3$.  These 
last two equations give two equivalent expressions for $B$ so by eliminating $B$ one gets a polynomial equation involving only $A$:
\bea
0&=A^4&\left[CxI_3+DJ_3-E(2xI_3+J_2)\right]\nonumber\\
&+A^3&\left[-CxI_1I_3+D(-3I_1J_3+2I_3J_1)+E(4xI_1I_3+3I_1J_2)\right]\nonumber\\
&+A^2&\left[C(-I_2I_3+I_3J_1^2)+D(-4I_1I_3J_1+3I_1^2J_3-I_3^2)
\right.\nonumber\\&&\left.
+E(-2xI_1^2I_3-3I_1^2J_2+2I_2I_3-I_3J_3)\right]\nonumber\\
&+A&\left[-C(I_1I_3J_1^2+I_3^2J_1)+D(I_1I_3^2+2I_1^2I_3J_1-I_1^3J_3)
\right.\nonumber\\&&\left.
+E(-2I_1I_2I_3+2I_1I_3J_3+I_1^3J_2-I_3^2J_1)\right]\nonumber\\
&&+E(I_1I_3^2J_1-I_1^2I_3J_3+I_3^3)~~,
\label{eq:q5a}
\eea
from whose solutions one can construct $B$ using
\be
B=-\frac{A^2D+A(J_1C-I_1D)-I_3E}{A(E-C)-I_1E}~~.
\label{eq:q5b}
\ee
The fourth order polynomial \Eq{eq:q5a}  has in general four solutions 
$A_1,\ldots,A_4$ at each point $x$ with an associated $B_1,\ldots,B_4$.  
The boundary conditions of perturbative behavior in the asymptotic spacelike 
region ($x\rightarrow\infty$), together with the criteria of continuity 
specify the physical solution.

\section*{Acknowledgments} 
\noindent
The authors would like to thank R.~Alkofer, M.~Bhagwat, M.~A.~Pichowsky and 
C.~D.~Roberts for useful discussions.  This work has been partially
supported by COSY (contract nos. 41139452, 41376610 and 41445395), 
and NSF grants no. PHY-0301190 and no. INT-0129236.  



\begin{thebibliography}{99}
\bibitem{Jain:1993qh}
P.~Jain, and H.~J.~Munczek, Phys. Rev. {\bf D48} (1993) 5403, [arXiv:hep-ph/9307221].
\bibitem{Maris:2003vk}
P.~Maris, and C.~D.~Roberts, Int. J. Mod. Phys. {\bf E12} (2003) 297, [arXiv:nucl-th/0301049].
\bibitem{Roberts:2000aa}
C.~D.~Roberts and S.~M.~Schmidt, Prog.\ Part.\ Nucl.\ Phys.\ {\bf 45} (2000) S1, [arXiv:nucl-th/0005064].
\bibitem{Maris:1999nt}
P.~Maris, and P.~C.~Tandy, Phys. Rev. {\bf C60} (1999) 055214, [arXiv:nucl-th/9905056].
\bibitem{Maris:2000sk}
P.~Maris, and P.~C.~Tandy, Phys. Rev. {\bf C62} (2000) 055204, [arXiv:nucl-th/0005015].
\bibitem{Volmer:2000ek}
J.~Volmer et al. (The Jefferson Lab F(pi) Collaboration), Phys. Rev. Lett. {\bf 86} (2001) 1713, [arXiv:nucl-ex/0010009].
\bibitem{Jarecke:2002xd}
D.~Jarecke, P.~Maris, and P.~C.~Tandy, Phys. Rev. {\bf C67} (2003) 035202, [arXiv:nucl-th/0208019].
\bibitem{Maris:2001am}
P.~Maris, PiN Newslett. {\bf 16} (2002) 213, [arXiv:nucl-th/0112022].
\bibitem{Maris:2002mz}
P.~Maris, and P.~C.~Tandy, Phys. Rev. {\bf C65} (2002) 045211, [arXiv:nucl-th/0201017].
\bibitem{Ji:2001pj}
C.-R.~Ji, and P.~Maris, Phys. Rev. {\bf D64} (2001) 014032, [arXiv:nucl-th/0102057].
\bibitem{Alkofer:2000wg}
R.~Alkofer, and L.~von Smekal, Phys. Rept. {\bf 353} (2001) 281, [arXiv:hep-ph/0007355].
\bibitem{Maris:1998hd}
P.~Maris, C.~D.~Roberts, and P.~C.~Tandy, Phys. Lett. {\bf B420} (1998) 267, [arXiv:nucl-th/9707003].
\bibitem{Munczek:1983dx}
H.~J.~Munczek, and A.~M.~Nemirovsky, Phys. Rev. {\bf D28} (1983) 181.
\bibitem{Bender:1996bb}
A.~Bender, C.~D.~Roberts, and L.~v.~Smekal, Phys. Lett. {\bf B380} (1996) 7, [arXiv:nucl-th/9602012].
\bibitem{Alkofer:2002bp}
R.~Alkofer, P.~Watson, and H.~Weigel, Phys. Rev. {\bf D65} (2002) 094026, [arXiv:hep-ph/0202053].
\bibitem{MarisPrivCom}
P.~Maris (2001), private communication.
\bibitem{Burden:1997nh}
C.~J.~Burden, L.~Qian, C.~D.~Roberts, P.~C.~Tandy, and M.~J.~Thomson, Phys. Rev. {\bf C55} (1997) 2649, [arXiv:nucl-th/9605027].
\bibitem{Burden:2002ps}
C.~J.~Burden, and M.~A.~Pichowsky, Few Body Syst. {\bf 32} (2002) 119, [arXiv:hep-ph/0206161].
\bibitem{Bloch:1999vk}
J.~C.~R.~Bloch, Y.~L.~Kalinovsky, C.~D.~Roberts, and S.~M.~Schmidt, Phys. Rev. {\bf D60} (1999) 111502, [arXiv:nucl-th/9906038].
\bibitem{Skullerud:2003qu}
J.~I.~Skullerud, P.~O.~Bowman, A.~Kizilersu, D.~B.~Leinweber, and A.~G.~Williams, JHEP {\bf 0304} (2003) 047, [arXiv:hep-ph/0303176].
\bibitem{Ball:1980ay}
J.~S.~Ball, and T.-W.~Chiu, Phys. Rev. {\bf D22} (1980) 2542.
\bibitem{Fischer:2003rp}
C.~S.~Fischer, and R.~Alkofer, Phys. Rev. {\bf D67} (2003) 094020, [arXiv:hep-ph/0301094].
\bibitem{Munczek:1995zz}
H.~J.~Munczek, Phys. Rev. {\bf D52} (1995) 4736, [arXiv:hep-th/9411239].
\bibitem{Bender:2002as}
A.~Bender, W.~Detmold, C.~D.~Roberts, and A.~W.~Thomas, Phys. Rev. {\bf C65} (2002) 065203, [arXiv:nucl-th/0202082].
\bibitem{Bhagwat:2004hn}
M.~S.~Bhagwat, A.~Holl, A.~Krassnigg, C.~D.~Roberts, and  P.~C.~Tandy (2004), [arXiv:nucl-th/0403012].
\bibitem{Maris:1997tm}
P.~Maris, and C.~D.~Roberts, Phys. Rev. {\bf C56} (1997) 3369, [arXiv:nucl-th/9708029].
\bibitem{Cornwall:1974vz}
J.~M.~Cornwall, R.~Jackiw, and E.~Tomboulis, Phys. Rev. {\bf D10} (1974) 2428.
\bibitem{Langfeld:2003ye}
K.\ Langfeld, H.\ Markum, R.\ Pullirsch, C.\,D.\ Roberts and S.\,M.\ 
Schmidt, Phys.\ Rev.\ {\bf C67} (2003) 065206, [arXiv:nucl-th/0301024].
\bibitem{Roberts:1996hh}
C.~D.~Roberts, Nucl. Phys. {\bf A605} (1996) 475, [arXiv:hep-ph/9408233].
\bibitem{Praschifka:1987pt}
J.~Praschifka, C.~D.~Roberts and  R.~T.~Cahill, Int. J. Mod. Phys. {\bf A2}
(1987) 1797.
\bibitem{Hollenberg:1992nj}
L.~C.~L.~Hollenberg, C.~D.~Roberts, and B.~H.~J.~McKellar, Phys. Rev.
{\bf C46} (1992) 2057.
\bibitem{Mitchell:1997dn}
K.~L.~Mitchell, and P.~C.~Tandy, Phys. Rev. {\bf C55} (1997) 1477, [arXiv:nucl-th/9607025].
\bibitem{Maris:2000ig}
P.~Maris, C.~D.~Roberts, S.~M.~Schmidt, and P.~C.~Tandy, Phys. Rev. {\bf C63} (2001) 025202, [arXiv:nucl-th/0001064].
\bibitem{Abramowitz:1968bk}
M.~Abramowitz, and I.~A.~Segun, "\textit{Handbook of Mathematical Functions}", Dover Publications (1968).
\end{thebibliography}
\end{document}